\documentstyle[12pt]{article}                                 

\begin{document}

\newcommand {\bea}{\begin{eqnarray}}
\newcommand {\eea}{\end{eqnarray}}
\newcommand {\be}{\begin{equation}}
\newcommand {\ee}{\end{equation}}

\title{
\begin{flushright}
\begin{small}
hep-th/9712118 \\
UPR-786-T\\
December 1997 \\
\end{small}
\end{flushright}
\vspace{1.cm}
Greybody Factors for Black Holes in Four Dimensions:\\
Particles with Spin}

\author{Mirjam Cveti\v{c} and Finn Larsen\\
\small Department of Physics and Astronomy\\
\small University of Pennsylvania\\
\small Philadelphia, PA 19104 \\
\small e-mail: cvetic,larsen@cvetic.hep.upenn.edu
}

\date{ }

\maketitle

\begin{abstract}
We compute the  emission spectrum of minimally coupled particles with spin
that are Hawking radiated from four dimensional black holes in string theory. 
For a  range of the black hole parameters  the result has 
a product structure that may be interpreted in terms of  the respective right-
and left-moving thermal correlation functions of 
an effective string model.
For spin-one  and spin-two particles a novel cancellation 
between contributions to the wave function is needed to ensure this outcome.  
The form of the spectra suggests that the four-dimensional effective 
string description is ``heterotic'': particles with spin are emitted
from the right-moving sector, only.
\end{abstract}                                

\newpage
\section{Introduction}
\label{sec:intro}
Recently a precise correspondence has been established between
black holes and collective states in string theory (for review 
see, {\it e.g.},~\cite{juanreview}). One of the results of these 
developments is that, in some cases, a classical black hole behaves 
as an effective string. The relevant low energy excitations 
of this effective 
description are the right- and left-moving modes of the string. 
A characteristic feature of this interpretation is that two
independent inverse temperatures $\beta_{R,L}$ can be introduced. 
Their physical significance
becomes apparent when Hawking radiation is interpreted microscopically 
as the result of colliding right- and left-moving excitations. In the
simplest case of minimally coupled scalar fields this yields
a emission spectrum with the characteristic factorized 
form~\cite{mathur,greybody}:
\be
\Gamma_{\rm em}(\omega)=
{P_R({\omega\over 2})\over e^{\beta_R\omega/2}-1}~
{P_L({\omega\over 2})\over e^{\beta_L\omega/2}-1}~{d^3 k\over (2\pi)^3}~,
\label {bosons}\ee
where $P_R$ and $P_L$ are polynomials in the frequency, with coefficients 
that depend on the  inverse temperatures $\beta_R$ and $\beta_L$, 
respectively.

The identification of the black hole with an effective string
faces a stringent test in the comparison of the emission spectrum
with the semi-classical result given by Hawking~\cite{hawking75}:
\be
\Gamma_{\rm em}(\omega) = \sigma_{\rm abs} (\omega) 
{1\over e^{\beta_H\omega}-1}~{d^3 k\over (2\pi)^3}~,
\label{eq:hawb}
\ee
where $\sigma_{\rm abs} (\omega)$ is the classical absorption 
cross-section for scalar particles impinging on the black hole and 
 $\beta_H={1\over 2}(\beta_R+\beta_L)$ is the
inverse Hawking temperature.  The agreement between the two descriptions
requires:
\be
\sigma_{\rm abs} (\omega) = {P_R({\omega\over 2})
\over e^{\beta_R\omega/2}-1}~
{P_L({\omega\over 2})\over e^{\beta_L\omega/2}-1}~(e^{\beta_H\omega}-1)~.
\label{eq:sigmaabs}
\ee
It is a striking confirmation of the effective string description
that the absorption cross-section indeed takes this form for a large range of
the black holes backgrounds and  energy ranges of the scattered particles.

The functions $P_{R,L}({\omega\over 2})$ have been calculated in the effective
string theory for minimally coupled scalars in the S-wave and in this case
a complete agreement with has been established that includes
the numerical coefficient~\cite{mathur,greybody}. 
Scalar particles with orbital angular momentum or non-minimal
couplings are  presently understood  with less precision,
because in these cases it is not known how to calculate fully the functions 
$P_{R,L}({\omega\over 2})$ in the effective string theory. However, 
qualitative arguments indicate that multiparticle interactions could
account for 
these processes~\cite{cgkt,mathur97,strominger97a,gubser,mathur97b,cl97b}.
It is therefore reasonable to take an absorption cross-section of the form 
eq.~\ref{eq:sigmaabs} as evidence for some underlying effective string 
description, even though the specifics of this theory remain unknown.

The purpose of the present paper is to calculate greybody factors for the
minimally coupled massless  
particles with spin-1/2, spin-1 and spin-2 in the background of static 
four-dimensional
  black holes that are parametrized by $4$ $U(1)$ charges (a prototype black
  hole solution of toroidally compactified string theory~\cite{cystrings,hlm}.)
.
The work is a contribution to  a broad program to address the  dynamical
properties of black holes  in string theory and thus  further shed light on the
microscopic structure of such black holes, in  particular to elucidate the
effective string interpretation.

The work also serves as a step toward a complete decoupling of the 
full set of perturbation equations for particle with different spins. 
We therefore  report on  partial results that are more general than are 
strictly needed here. Other recent work on greybody factors for 
particles with spin can be found in~\cite{gibbons96,sdas,gubser97b,hoso}. 

Let us summarize the main results. In the simplest case of 
massless (minimally coupled)  Weyl fermions the  absorption
cross-section takes the following form:
\be
\sigma_{\rm abs} (\omega) = {P_R({\omega\over 2})
\over e^{\beta_R\omega/2}+1}~
{P_L({\omega\over 2})\over e^{\beta_L\omega/2}-1}~(e^{\beta_H\omega}+1)~,
\label{eq:sigmaabsf}
\ee
where again $P_R$ and $P_L$ are polynomials in the frequency that
depend on $\beta_R$ and $\beta_L$, respectively. 
 A striking feature of this expression is the  Fermi-Dirac factor
$e^{\beta_H\omega}+1$ in the numerator, which  precisely
cancels the Fermi-Dirac factor in Hawking's expression for the emission rate:
\be
\Gamma_{\rm em}(\omega) = \sigma_{\rm abs} (\omega) 
{1\over e^{\beta_H\omega}+1}~{d^3 k\over (2\pi)^3}~,
\label{eq:hawf}
\ee
and thus yields an  emission rate that is compatible with an interpretation 
as a correlation function in a weakly coupled effective string theory. 
In this description of emission the statistical factors in the denominator 
of eq.~\ref{eq:sigmaabsf}  have their origin as  phase space factors of
the right- and left-moving string states. 
The fact that the right-moving  factor is of the  Fermi-Dirac type while
the left factor is of  the Bose-Einstein type indicates that the
effective string theory for the four dimensional black hole
is ``heterotic'', i.e.  the right-moving sector has both fermionic and bosonic
degrees of freedom, while
the left-moving sector has only   bosonic degrees of freedom. 
It has previously been observed that the entropy formula for rotating black
holes in four dimensions suggests that the angular momentum is carried only
by the right-moving modes~\cite{cy96b,strominger97a}. 
This is in  harmony with our result because it is exactly the worldsheet
fermions that carry the spacetime angular momentum~\cite{rotation1}.

For minimally coupled spin-1 and spin-2 fields we again find absorption 
cross-sections of the factorized form eq.~\ref{eq:sigmaabs}. The 
factorization turns out to be nontrivial for particles with spin $s\geq 1$: 
in intermediate steps of the calculation there is a polynomial 
dependence on $\beta_H={1\over 2}(\beta_R+\beta_L)$. However, this 
dependence cancels in the final expression for emission rates and so our
results provide a nontrivial test of the effective string model.

The emission rate for  particles with arbitrary spin and total 
angular momentum
can be written in a concise form that contains fermionic and
bosonic degrees of freedom on equal footing. This result suggests that the 
microscopic theory is supersymmetric even though we consider black hole 
backgrounds that are not necessarily extremal. The regularity that we see 
may be a manifestation of the supersymmetry that is present in the theory,
but broken by the presence of the black hole. Our result 
indicates that for black holes in theories with supersymmetry there is 
a  relation between the  absorption cross-sections  for particles with
different spins. This would seem to suggest that the effective string 
is a superstring with supersymmetry broken only by the effects
of thermodynamics.

The paper is organized as follows: in sec.~\ref{sec:bh}
we summarize the classical geometry of the black holes backgrounds
we consider, and in section~\ref{sec:np} we describe these black holes in the
Newman-Penrose formalism. This sets the stage for the derivation
of field equations for perturbations, in sec.~\ref{sec:eom}.
For spin-1/2 and spin-1 we consider ``spectator'' particles that
respond to the gravitational field but not to the background gauge 
fields, and are thus minimally coupled. In the case of gravitons we 
are not able to decouple the 
general perturbation equations but we can do so in an approximation 
that is sufficient to establish absorption cross-sections of the
effective string form. 
In sec.~\ref{sec:flux} we comment on the features of the
field equations, and find the general relation between their solutions and
the absorption cross-section. This is exploited in sec.~\ref{sec:abs}, 
where approximate wave functions are calculated. In sec.~\ref{sec:results} 
we discuss the final results and consider some special cases and in 
sec.~\ref{sec:microscopics} we  discuss their relation  
to the effective string theory model. Finally, in sec.~\ref{sec:conclusion}, 
we conclude with  comments on the generality of the type of greybody 
factors considered here.

\section{The Black Hole Geometry}
\label{sec:bh}
We consider the black holes in string theory that 
are characterized by
their mass $M$ and four $U(1)$ charges $Q_i$~{\cite{cystrings,hlm}}. 
These quantum numbers are parametrized as:
\bea
M &=& {1\over 2}\mu\sum_{i=1}^4 \cosh 2\delta_i~, \\
Q_i &=& {1\over 2}\mu \sinh 2\delta_i~~;~~~i=1,2,3,4~.
\eea
The gravitational coupling constant in four dimensions is $G_4={1\over 8}$. 
In string theory this corresponds to the relation 
$g^2(2\pi)^6(\alpha^\prime)^4/V_6=1$ between the gauge coupling $g$, 
the string tension $\alpha^\prime$ and the volume of the six-dimensional 
compactified space $V_6$.\footnote{The $m$ of~\cite{cy96b} 
is $m={1\over 4}\mu$ and the $r_0$ of~\cite{hlm} is $r_0={1\over 2}\mu$.}
The black hole metric can be written:
\be
ds^2 = - {\Delta\over\Sigma}dt^2 + \Sigma ({1\over\Delta}dr^2
+ d\theta^2+\sin^2\theta d\phi^2)~,
\label{eq:metric}
\ee
where:
\bea
\Delta &=& r^2 - {1\over 16}\mu^2~,
\label{eq:Delta}\\
\Sigma &=& \prod_i (r+{1\over 4}\mu\cosh 2\delta_i)^{1\over 2}~.
\label{eq:Sigma}
\eea
Note that in these coordinates the two horizons are located at:
\be
r_{\pm} =\pm{1\over 4}\mu~.
\ee

The black hole can be interpreted as a generalization of the standard
Reissner-Nordstr\"{o}m solution. This special case is obtained by 
identifying the four charges through:
\bea
2Q_{RN}&=& Q_1 = Q_2 = Q_3 = Q_4 = {1\over 2}\mu\sinh 2\delta~, \\
M_{RN} &=& 2\mu\cosh 2\delta~,
\eea
and transforming our radial coordinate to the conventional one:
\be
r_{RN} = r + {1\over 4}\mu\cosh\delta~.
\ee
Then eqs.~\ref{eq:Delta}-\ref{eq:Sigma} become:
\bea
\Sigma_{RN} &=& r^2_{RN}~,\\
\Delta_{RN} &=& r_{RN}^2 - 2 ~~ {1\over 8}Mr_{RN} + Q^2_{RN}~,
\eea
and the metric~eq.~\ref{eq:metric} indeed reproduces the  
Reissner-Nordstr\"{o}m line element with $G_N = {1\over 8}$.

The entropy and the inverse temperature of the general black hole are:
\bea
S  &\equiv & {A\over 4G_N}=2\pi\mu^2\prod_i \cosh\delta_i~,  \\
\beta_H &\equiv & {2\pi\over\kappa_{+}} = 2\pi\mu\prod\cosh\delta_i~,
\eea
where $A$ and $\kappa_{+}$ denote the area and the surface
acceleration, both measured at the outer horizon. We will also need 
the right (R) and left (L) temperatures that combine  the respective 
surface accelerations  $\kappa_{\pm}$ 
at the outer and inner horizons~\cite{fl97}:
\bea
\beta_R &=& {2\pi\over\kappa_{+}}+{2\pi\over\kappa_{-}} 
=2\pi\mu (\prod_i \cosh\delta_i+\prod_i\sinh\delta_i)~,
\\
\beta_L &=& {2\pi\over\kappa_{+}}-{2\pi\over\kappa_{-}} =
2\pi\mu (\prod_i \cosh\delta_i-\prod_i \sinh\delta_i)~.
\eea

The metric is a solution to the equations of motion low energy
string theory in the
presence of matter that can be represented as a collection of $U(1)$ 
gauge fields and scalars. For the study greybody factors  we will not 
need the explicit form of these matter fields. This state of affairs 
points to a  robustness of this kind of calculation: these black 
holes can be considered solutions to different theories. 
It is perhaps natural to consider the low energy limits of 
toroidally compactified string theories, namely the $N=4$ or $N=8$ 
supergravity theories. However, the full family of solutions is still allowed 
in the bosonic part of the S-T-U--model of $N=2$ supergravity
and they are therefore also solutions to $N=2$ supergravity. 
Thus, the greybody factors of the type  found in this paper are generic 
for black holes that can be embedded in $N=4,8$ supergravity theory,
even when this possibility is not realized.

\section{The Newman-Penrose Formalism}
\label{sec:np}
The Newman--Penrose formalism greatly simplifies 
the consideration of particles with spin in curved spacetime.
In the following we summarize the basic features
and define the notation. Here and throughout the paper we rely 
heavily on the book by Chandrasekhar~\cite{chandra}. For other 
introductions to the Newman--Penrose formalism we refer 
to~\cite{wald,rindler}. Some relevant discussion and applications 
were given in~\cite{gubser97b}.

The starting point is the choice of a complex null-tetrad. We take:
\bea
{\bf l} &=& -dt + {\Sigma\over\Delta}dr~, \\
{\bf n} &=& -{\Delta\over 2\Sigma}dt - {1\over 2}dr~, \\
{\bf m} &=& \sqrt{\Sigma\over 2} (d\theta + i\sin\theta d\phi)~.
\eea
The dual basis is of the form:
\be
l^\mu \partial_\mu = \partial_r + {\Sigma\over\Delta}\partial_t~~~;~~
n^\mu \partial_\mu = -{\Delta\over 2\Sigma}\partial_r + 
{1\over 2}\partial_t~~;~~~
m^\mu \partial_\mu = {1\over\sqrt{2\Sigma}}(\partial_\theta - 
{i\over\sin\theta}\partial_\phi)~.
\ee 
These directional partial derivatives are
denoted $(D,\Delta,\delta)$ in the Newman-Penrose literature but 
we will not do so, to avoid confusion with other meanings of those 
symbols. The tetrad is normalized so that $l\cdot n=-1$ 
and $m\cdot {\bar m}=1$. With our choice of basis the
vectors $n^\mu$ and $l^\mu$ parametrize infalling and outgoing 
null-geodesics, with the condition that the latter geodesic is 
affinely parametrized.

In the local frame a vector index is equivalent to two spinor indices. 
The Clebsch-Gordon coefficients $\sigma^\mu_{AA^\prime}$ needed for 
this translation are conventionally suppressed. Thus the covariant
directional derivatives can be written:
\be
\nabla_{\hat l} = \nabla_{00^\prime}
~~;~~
\nabla_{\hat n} = \nabla_{11^\prime}
~~;~~
\nabla_{\hat m} = \nabla_{01^\prime}
~~;~~
\nabla_{\hat {\bar m}} = \nabla_{10^\prime}~.
\ee
Tensors are always symmetric in primed and unprimed spinor indices
independently, but there is no special symmetry relating the
two kinds of indices. Spinor indices are raised and lowered 
using the convention
$\chi^0=\epsilon^{01}\chi_1$ where $\epsilon^{01}=1=-\epsilon^{10}$.
 
In the Newman--Penrose formalism the components 
of the spin-connection, needed for the evaluation of covariant
derivatives, are referred to as the spin--coefficients. The 12 complex
spin-coefficients have conventional names given as:
\be
\begin{array}{ccc}
\gamma_{00^\prime 00}=\kappa &
\gamma_{00^\prime 10}=\epsilon &
\gamma_{00^\prime 11}=\pi \\
\gamma_{10^\prime 00}=\rho &
\gamma_{10^\prime 10}=\alpha &
\gamma_{10^\prime 11}=\lambda \\
\gamma_{01^\prime 00}=\sigma &
\gamma_{01^\prime 10}=\beta &
\gamma_{01^\prime 11}=\mu \\
\gamma_{11^\prime 00}=\tau &
\gamma_{11^\prime 10}=\gamma &
\gamma_{11^\prime 11}=\nu 
\end{array}
\ee
The most efficient way to calculate the spin-coefficients is to use 
Cartan's structure equations. By explicit calculation we find:
\bea
d{\bf l} & = & 0~,\\
d{\bf n} & = & \partial_r {\Delta\over 2\Sigma}~{\bf l}\wedge {\bf n}~,\\
d{\bf m} & = & {\Delta\partial_r\Sigma\over 4\Sigma^2}~
{\bf l}\wedge {\bf m}-{\partial_r\Sigma\over 2\Sigma}~
{\bf n}\wedge {\bf m}
+{\cot\theta\over\sqrt{2\Sigma}} {\bar{\bf m}}\wedge{\bf m}~. 
\eea
Comparison with the standard form of the structure equations
gives the nonvanishing spin-coefficients:
\bea
\beta & = & {\cot\theta\over 2\sqrt{2\Sigma}} = -\alpha~,\\
\gamma & = & \partial_r {\Delta\over 4\Sigma}~,\\
\mu  & = & -{\Delta\partial_r\Sigma\over 4\Sigma^2}~,\\
\rho & = & - {\partial_r\Sigma\over 2\Sigma}~.
\eea
The spin-coefficients provide the information that is needed to
evaluate covariant derivatives or, in physical terms, 
to translate polarization vectors along geodesics. 

The Weyl tensor is the irreducible part of the Riemann curvature 
tensor after the Ricci tensor has been projected out. 
In four dimensions 
it has $10$ independent components that are represented as $5$ complex 
numbers $\Psi_i$ in the Newman-Penrose formalism. For
the specific metric we consider most components vanish:
\be
\Psi_0=\Psi_1=\Psi_3=\Psi_4=0~.
\label{eq:typeD}
\ee
This property is the definition of a spacetime that is of type D
in the Petrov classification. The last component
of the Weyl tensor is non-trivial: 
\be
\Psi_2 = -\partial_r\Delta {\partial\Sigma\over 4\Sigma^2}
-{\Delta\over 3}({\partial^2 \Sigma\over 2\Sigma^2} - 
{(\partial\Sigma)^2\over \Sigma^3})~.
\ee
In its simplest form the Goldberg-Sachs theorem states that the 
type D property is equivalent to:
\be
\kappa = \sigma = \lambda = \nu = 0~,
\label{eq:typeD2}
\ee
in vacuum spacetimes. This version of the theorem is not applicable 
in our case, because matter is present. However, the spacetime
we consider nevertheless has both simplifying features, 
eqs.~\ref{eq:typeD} and~\ref{eq:typeD2}. 
This is important for the consideration of perturbations.

The $10$ components of the Ricci tensor are represented in the
Newman-Penrose formalism as a spinorial 
tensor $\Phi_{ABA^\prime B^\prime}$ with $9$ independent 
components and the scalar $\Lambda$ that is proportional to the Ricci 
scalar. In a spherically symmetric spacetime it is only the real 
components that do not vanish:
\bea
\Phi_{000^\prime 0^\prime} &=& -{\partial^2_r\Sigma\over 2\Sigma} + 
{(\partial_r\Sigma)^2\over 4\Sigma^2}~,
\label{eq:ricci1} \\
\Phi_{010^\prime 1^\prime} &=&{1\over 2\Sigma}-\partial_r\Delta 
{\partial_r\Sigma\over 4\Sigma^2}
-{\Delta\over 4}({\partial^2_r\Sigma\over 2\Sigma^2} - 
{3(\partial_r\Sigma)^2\over 4\Sigma^3})~,\\
\Phi_{111^\prime 1^\prime} &=& -{\Delta^2\over 4\Sigma^2}
({\partial^2_r\Sigma\over 2\Sigma} -
{(\partial_r\Sigma)^2\over 4\Sigma^2})~,\\
\Lambda &=& - {\Delta\over 12}( {\partial^2_r\Sigma\over 2\Sigma^2}
-{(\partial_r\Sigma)^2\over 4\Sigma^2})~.
\label{eq:ricci4}
\eea
In deriving these formula the special form of $\Delta$ has been
exploited to write $\partial_r^2\Delta=2$, but $\Sigma$ has been 
kept arbitrary.

\section{Field Equations for Perturbations}
\label{sec:eom}
In the present article we take the field equation:
\be
\nabla_{AA^\prime}\psi^{AB_1\cdots B_{2s-1}} = 0~,
\label{eq:fieldeq}
\ee
as the starting point for the discussion. This is the natural
covariantization of the field equation for spin $s$ in flat space. 
For spin $s=2$ the equation is inadequate as it stands and 
additional non-linear terms must be added for consistency. We will discuss 
the required modification for $s=2$ later in this section.

For $s={1\over 2}$ and $s=1$ the field equation is consistent but the theory
may not contain fields that couple to the background in this way. A case 
in point is the $N=8$ supergravity. In this case the 
linearized field equations for the fermions with spin $s={1\over 2}$ are (see, 
{\it e.g.},~\cite{cremmer}):
\be
\nabla_{AA^\prime}\psi^A_{ijk} + 
\epsilon_{ijklmnop}F^{lm}_{A^\prime B^\prime }\psi^{B^\prime nop}
 = 0~,
\label{eq:genferm}
\ee
where the $SO(8)$ indices have been denoted by small roman letters. (We have
also suppressed terms proportional to the gravitino perturbations.)
The  four background $U(1)$ fields can be 
represented after diagonalization as $F^{12}$, $F^{34}$, $F^{56}$, 
and $F^{78}$.  It is therefore apparent that in this case the
fermions of $N=8$ supergravity generically couple to the background 
gauge fields.

A similar situation is found for the vector particles in $N=8$
supergravity. They satisfy the linearized field equations of the type:
\be
\nabla_{AA^\prime} (\delta_{ij}^{kl} + iM_{ij}^{kl})F_{kl}^{A^\prime B}=0~,
\label{eq:genvec}
\ee
where the $M_{ij}^{kl}$ parametrize the scalar moduli fields. 
The vector
particles in $N=8$ supergravity therefore generically couple to the 
scalar particles. 

In the present work we only take into account the 
gravitational couplings of the fermions and vectors. The philosophy 
is similar to that of considering {\it minimally coupled} scalars in
general relativity: it is the simplest couplings that display
the effects  in the scattering phenomena.  Naturally it would be 
interesting to investigate 
the effect of background $U(1)$ charges and scalars in the context of, 
{\it e.g.}, $N=8$ supergravity.

Let us now return to the field equations. In
component form they are:
\bea
&-&k\sigma\psi^{0\cdots 0\overbrace{1\cdots 1}^{k-1}} 
+[m^\mu \partial_\mu + 2(s-k)\beta - 
(k+1)\tau]\psi^{0\cdots 0\overbrace{1\cdots 1}^{k}} +\\ 
&+&[n^\mu \partial_\mu - 2(s-k-1)\gamma +(2s-k)\mu]
\psi^{0\cdots 0\overbrace{1\cdots 1}^{k+1}} 
+(2s-k-1)\nu\psi^{0\cdots 0\overbrace{1\cdots 1}^{k+2}}  
= 0 \nonumber \\
&-&k\kappa\psi^{0\cdots 0\overbrace{1\cdots 1}^{k-1}} 
+[l^\mu \partial_\mu + 2(s-k)\epsilon -(k+1)\rho]
\psi^{0\cdots 0\overbrace{1\cdots 1}^{k}} +\\
&+&[{\bar m}^\mu \partial_\mu -2(s-k-1)\alpha +(2s-k)\pi]
\psi^{0\cdots 0\overbrace{1\cdots 1}^{k+1}} 
+(2s-k-1)\lambda\psi^{0\cdots 0\overbrace{1\cdots 1}^{k+2}}  
= 0  \nonumber
\eea
for $k=0,\cdots,2s-1$. Each equation is a relation between only two 
components of the wave function, when $\kappa=\sigma=\lambda=\nu$.
In this case each pair of equations in fact determines the two 
components of the wave function. In this way we recognize 
the central role of the type-D property for the decoupling of 
perturbation equations.

At this point we use the explicit expressions for the spin-coefficients
and find:
\bea
{\cal L}_{s-k}\tilde{\psi}^{0\cdots 0\overbrace{1\cdots 1}^{k}}
-{\Delta}^{1/2}({\cal D}^\dagger_{{k+1\over 2}-s} - {2k+1-2s\over 2}
{\partial_r\Sigma\over\Sigma}) 
\tilde{\psi}^{0\cdots 0\overbrace{1\cdots 1}^{k+1}} &=& 0 \\ 
{\cal L}^\dagger_{k+1-s}\tilde{\psi}^{\overbrace{1\cdots 1}^{k+1}}
-{\Delta}^{1/2}({\cal D}_{k/2} + {2k+1-2s\over 2}
{\partial_r\Sigma\over\Sigma}) 
\tilde{\psi}^{0\cdots 0\overbrace{1\cdots 1}^k } &=& 0
\eea
where:
\bea
{\cal D}_n &=& \partial_r - {i\omega\Sigma\over\Delta} + 
n{\partial_r\Delta\over\Delta}~,\\
{\cal D}^\dagger_n &=& \partial_r + {i\omega\Sigma\over\Delta} + 
n{\partial_r\Delta\over\Delta}~, \\
{\cal L}_n &=& \partial_\theta + {m\over\sin\theta} + n\cot\theta~, \\
{\cal L}_n^\dagger &=& \partial_\theta - {m\over\sin\theta} + n\cot\theta~.
\eea
We also introduced a rescaled wave function ${\tilde\psi}$ through:
\be
\psi^{0\cdots 0\overbrace{1\cdots 1}^{k}} = 
\Delta^{-k/2}
(2\Sigma)^{k/2-s}~\tilde{\psi}^{0\cdots 0\overbrace{1\cdots 1}^{k}}~.
\ee
The equations take their most symmetric form when written in terms of 
$\tilde{\psi}$. We can separate variables by writing:
\be
\psi^{0\cdots 0\overbrace{1\cdots 1}^{k}} = \Delta^{-k/2}
(2\Sigma)^{k/2-s}P_{k-s}(r)S_{k-s}(\Omega)~.
\label{eq:radial}
\ee
The angular functions $S_{k-s}$ satisfy:
\bea
{\cal L}^\dagger_{k+1-s}{\cal L}_{s-k}S_{k-s} &=& -
\Lambda^{(-)}_{k-s} S_{k-s}~,\\
{\cal L}_{s-k}{\cal L}^\dagger_{k+1-s}S_{k+1-s} &=& 
-\Lambda^{(+)}_{k+1-s} S_{k+1-s}~,
\eea
where the $\Lambda^{(\pm )}_k$ are separation constants.
The radial function $P_{k-s}$ similarly satisfy:
\bea
\Delta^{1/2} [{\cal D}^\dagger_{{k+1\over 2}-s}-(2k+1-2s)
{\partial\Sigma\over 2\Sigma} ]
\Delta^{1/2} [{\cal D}_{-{k\over 2}}+(2k+1-2s)
{\partial\Sigma\over 2\Sigma} ]
P_{k-s} \nonumber = \\ = \Lambda^{(-)}_{k-s} P_{k-s} \\
\Delta^{1/2} [{\cal D}_{-{k\over 2}}+(2k+1-2s)
{\partial\Sigma\over 2\Sigma} ]
\Delta^{1/2} [{\cal D}^\dagger_{{k+1\over 2}-s}-(2k+1-2s)
{\partial\Sigma\over 2\Sigma} ]
P_{k+1-s} = \nonumber \\ = \Lambda^{(+)}_{k+1-s} P_{k+1-s} 
\eea
There is one equation for each of the upper and lower components 
of the wave function $P_s$ and $P_{-s}$. For the remaining components 
$P_{\lambda}; \lambda= -s+1,\cdots,s-1$ there are two equations that 
are in general distinct. The functions are therefore overdetermined 
and it is only because of the underlying algebraic structure that there 
are solutions at all. Similar comments apply to the angular equations.

The angular functions embody the properties of the rotation group. 
We denote the total angular momentum of the particle $j$, its projection 
on a fixed axis $m$, and the helicity 
$\pm s$.\footnote{The total angular momentum is often denoted $l$.
Our notation is intended to avoid confusion with the orbital angular 
angular momentum that plays no role in the present computation.}
The angular momentum satisfies $j\ge s$. The separation constants 
$\Lambda^{(\pm )}_{k-s}$ can be determined by algebraic or analytical 
methods, with the result: 
\be
\Lambda^{(\pm)}_{k-s} = j(j+1) - (k-s)(k-s\mp 1)~.
\ee
We will only consider the highest and lowest component of the wave 
function. Each is associated with a single separation constant, 
and their respective values coincide:
\be
\Lambda \equiv \Lambda^{(-)}_{-s} = \Lambda^{(+)}_s = j(j+1) - s(s-1)~.
\ee
The explicit angular wave functions are known. They are closely related 
to the Jacobi-Polynomials and they are proportional to the rotation 
matrices:
\be
S_k (\Omega ) = e^{im\phi}d^{(j)}_{km} (\theta)~,
\ee
where $k=-s,\cdots,s$.

Our primary interest is the radial equation for the upper and lower
components of the wave function. The upper one can be written:
\bea
&~&\left\{ \Delta^s \partial_r \Delta^{1-s}\partial_r + 
{\Sigma^2\omega^2 -is\omega\Sigma\partial_r\Delta\over\Delta }
+ 2is\omega\partial_r\Sigma + \right. \nonumber \\
&+&\left. \Delta (s-{1\over 2})[\partial_r ({\partial_r\Sigma\over\Sigma})
+ (s-{1\over 2}) ({\partial_r\Sigma\over\Sigma})^2 + 
(1-s){\partial_r\Delta\over\Delta}~{\partial_r\Sigma\over\Sigma}]
\right\} P_s = \Lambda P_s
\label{eq:rawmaster}
\eea
and the lower one is the complex conjugate. An equation equivalent to 
this one was found by Gubser~\cite{gubser97b}. 

We now consider gravitons in more detail. In this case the full 
perturbation equations involve all the components of the curvature 
tensor, a formidable problem. However, in type D spacetimes the 
radiative parts of the field decouple and form a manageable subset. 
Namely, we consider perturbations of the $(\Psi_0,\Psi_1,\Psi_3,\Psi_4)$ 
components of the Weyl tensor and identify them with the 
fields $(\psi^{1111},-\psi^{0111},-\psi^{0001},\psi^{0000})$
in the general calculation. The Weyl tensor is related to the
covariant derivatives of the spin--coefficients:
\bea
\Psi_0 &=& 
(l^\mu \partial_\mu - \rho - \rho^\star - 3\epsilon + \epsilon)\sigma
-(m^\mu\partial_\mu - \tau +\pi^\star - \alpha^\star - 3\beta)\kappa
\label{eq:psi0}
\\
\Psi_4 &=& 
(n^\mu \partial_\mu + \mu+\mu^\star + 3\gamma - \gamma^\star)\lambda
-({\bar m}^\mu \partial_\mu + 3\alpha + \beta^\star + \pi -\tau^\star)\nu
\label{eq:psi4}
\eea
Perturbations in the Weyl tensor are therefore necessarily accompanied
by perturbations in the spin--coefficients $(\kappa,\sigma,\nu,\lambda)$. 
In the spin-2 case such terms should therefore be kept in the derivation
above. They always appear multiplied by the non-vanishing component of 
the Weyl-tensor, namely $\Psi_2$. 

There is an additional complication that must be taken into 
account: in the spin-2 case the rationale for the field equation 
eq.~\ref{eq:fieldeq} is its relation to the Bianchi identity.
In vacuum it is indeed exactly the Bianchi identity but in the 
presence of matter there are additional couplings between 
$(\kappa,\sigma,\nu,\lambda)$ and the Ricci tensor given in 
eqs.~\ref{eq:ricci1}-\ref{eq:ricci4}. The components 
$\Phi_{000^\prime 0^\prime}$ and $\Phi_{111^\prime 1^\prime}$ vanish in the
Reissner-Nordstr\"{o}m case but in the general case they lead to
couplings between $(\kappa,\sigma,\nu,\lambda)$ and the
complex conjugate fields 
$(\kappa^\star,\sigma^\star,\nu^\star,\lambda^\star)$.
This is reminiscent of the Pauli couplings in 
eqs.~\ref{eq:genferm}-\ref{eq:genvec}.

The net effect of all this is that we can justify eq.~\ref{eq:rawmaster} 
for gravitons, with the amendment that certain source terms that are 
proportional to $(\kappa,\sigma,\nu,\lambda)$ and their complex
conjugates must be added. In the Reissner-Nordstr\"{o}m limit the 
combinations of spin--coefficients that appear in these source terms 
are exactly such that they can be eliminated, using the Ricci-identities 
eqs.~\ref{eq:psi0}-\ref{eq:psi4}. Moreover, the 
resulting term proportional to $\Psi_4$ precisely cancel the term
in the square bracket of eq.~\ref{eq:rawmaster}~\cite{chandra}. 
For general $U(1)$ charges we are not able to decouple the 
field equations in a similar way and we are left with the source terms. 
However,
a calculation shows that, in the approximation that we will use in 
sec.~\ref{sec:abs}, most source terms are negligible: they  
are multiplied by an explicit factor of $\Delta$ that make them vanish 
close to the horizon. Moreover, the cancellations noted in the 
Reissner-Nordstr\"{o}m case are still approximately valid and ensure 
that source terms fall off at large distances at a rate faster than the 
angular momentum term $\Lambda$. In the following we will use
use eq.~\ref{eq:rawmaster} with the term in the brackets omitted as an 
approximate decoupled field equation for gravitons. This procedure
is exact in the Reissner-Nordst\"{o}m case and it should give a reliable 
indication in general.

The case of gravitini, { i.e.} $s={3\over 2}$, is similar to that
of gravitons: eq.~\ref{eq:rawmaster} should give some guidance but the 
full equations may contain additional terms. It is possible that 
the correct equations can be found by omitting the term in square 
bracket, but we have made no effort to substantiate this speculation
for $s={3\over 2}$. However, the square bracket vanishes for $s={1\over 2}$; 
and for $s=1$ it is simple to verify that it will not 
contribute within our approximation scheme. We will therefore omit this 
term in general and take the approximate equation:
\be
[\Delta^s \partial_r \Delta^{1-s}\partial_r + 
{\Sigma^2\omega^2 -is\omega\Sigma\partial_r\Delta\over\Delta}
+ 2is\omega\partial_r\Sigma]P_s = \Lambda P_s~,
\label{eq:master}
\ee
as a basis for further exploration. This decoupled radial equation 
applies also for spin-0, as can be verified by inspection. 
An equation of this type was first found by Teukolsky, in the case of the 
neutral Kerr black hole background, and so it is sometimes
referred to as the Teukolsky equation~\cite{teukolsky}.

\section{The Flux Factors}
\label{sec:flux}
The Teukolsky equation eq.~\ref{eq:master} is satisfied by the upper
component of the wave function. The lower component satisfies a
different differential equation, namely the complex conjugate one.
In order to solve the scattering problem we must find solutions 
to each of the two equations and, importantly, we must find a
relation between the two components of the wave function
that have thus been identified. The solution to this problem is 
straightforward in the spin-1/2 case: the original first 
order differential equation essentially gives one component as the 
derivative of the other one.
Similar results are valid in the higher spin cases where the needed
relations are known as the Teukolsky-Starobinsky identities. In general
it requires a comprehensive analysis of the complete system of first 
order equations to find this result, but for our purpose the following 
approximate procedure is sufficient: at the horizon the two linearly 
independent solutions to the equation for the upper component 
of the wave function are an infalling wave and a solution that vanishes. 
The corresponding solutions for the lower component of the wave function 
are the complex conjugate ones.
In the absorption geometry we consider the solution with ingoing flux 
at the horizon. Using the differential equation for the upper component 
we find the associated amplitude at infinity. This cannot be the 
whole story because the flux at the horizon and at infinity are not the 
same and so there must also be an outgoing flux at infinity, 
corresponding to the reflected wave. 
This flux is carried by the lower component of the wave function 
and the magnitude can be determined from flux conservation. In the cases 
we consider the transmission is in fact small and so the reflected flux 
is identical to the incoming one, up to quantities of subleading order.
It is therefore sufficient to consider only the upper component
of the wave function.

The discussion in the previous paragraph applies to a particle in a 
specific helicity state. The corresponding treatment of particles with 
the opposite helicity simply involves complex conjugation and the 
interchange of upper and
lower components of the wave function. For this reason the two helicities
lead to the same absorption cross-section, as expected for scattering
off a parity invariant target. We consider the particles with
left-handed chirality, for definiteness.

We now turn to the main task of this section, to determine the flux 
factors that are needed to convert the upper component of a
wave function 
to an absorption cross-section. The general result:
\be
\sigma_{\rm abs}(\omega) = {\pi\over\omega}(2j+1) |T|^2~,
\ee
reduces the problem to one of transmission in one spatial dimension. 
We normalize the wave functions at infinity 
as:
\be
P_s^{(\infty)} \sim A_s^{(\infty)} (2r\omega)^{2s-1}~e^{-i\omega r}~.
\ee
Similarly, close to the outer horizon of the black hole:
\be
P_s^{(0)} \sim A_s^{(0)} (r-r_{+})^{-i{\beta_H\omega\over 4\pi}}~.
\ee
After this general statement of the problem we consider each case 
independently.

\paragraph{spin  1/2 
:}
The conserved current is:
\be
{1\over\sqrt{2}}J^\mu = \sigma^\mu_{AB^\prime}\psi^A {\bar\psi}^{B^\prime}~,
\ee
where:
\be
\sigma^\mu_{AB^\prime} = {1\over\sqrt{2}}
\left( \begin{array}{cc}
l^\mu & m^\mu \\
{\bar m}^\mu & n^\mu 
\end{array} \right)~.
\ee
Recalling the definition of the radial wave functions
eq.~\ref{eq:radial} the radial current becomes:
\be
J^r = - {1\over 2\Sigma} 
(|P_{1/2}S_{1/2}|^2 - 
|P_{-{1/2}}S_{-{1/2}}|^2 )~.
\ee
We normalize the angular wave functions:
\be
{1\over 4\pi}\int |S_{\pm s}|^2 d\Omega = 1~,
\ee
and so the final result for the infalling flux becomes:
\be
{1\over 2\pi}{dN\over dt} = - {1\over 2\pi}
\int J^r~\Sigma\sin\theta d\theta d\phi =
|P_{1/2}|^2-|P_{-{1/2}}|^2~.
\ee
This formula can be used both at the horizon and at infinity. The
effective two-dimensional transmission coefficient is therefore simply:
\be
|T_{s={1/2}}|^2 =  | { A^{(0)}_{1/2}\over 
A^{(\infty)}_{1/2} }|^2 ~.
\ee

\paragraph{spin 1:}
For spin $s>{1\over 2}$ there are no conserved currents. However,
the flux factors can be inferred from the flow of energy.
For a spin-1 field the energy momentum tensor is:
\be
T^{\mu\nu} = 2\sigma^\mu_{AA^\prime}\sigma^\nu_{BB^\prime}
\psi^{AB}{\bar\psi}^{A^\prime B^\prime}~.
\ee
This works out to:
\be
T^{rt} = - {1\over 4\Sigma\Delta}
(|P_1 S_1 |^2 - 
|P_{-1}S_{-1}|^2 )~.
\ee
The local energy of photons is corrected for redshift according to:
\be
E = {\Sigma\over\Delta}E_\infty =  {\Sigma\omega\over\Delta}~,
\ee
and so the inflowing flux becomes:
\be 
{1\over 2\pi}~
{dN\over dt} = -{1\over 2\pi}\int {1\over E}T^{rt}~\Sigma d\Omega
=  -{1\over 2\pi\omega}\int \Delta T^{rt} d\Omega
= {1\over 2\Sigma\omega} (|P_1|^2 - |P_{-1}|^2 )~.
\ee
In the vicinity of the horizon $\Sigma\simeq {1\over 8\pi}\mu\beta_H$
while at infinity $\Sigma\sim r^2$. The effective transmission 
coefficient becomes:
\be
|T_{s=1}|^2 =  {2\pi\over\mu\beta_H}| { A^{(0)}_1\over 
A^{(\infty)}_1 }|^2 ~.
\ee

\paragraph{spin 2:}
In the spin two case there is in general no conserved energy momentum 
tensor. For this and related reasons it is somewhat ambiguous to refer 
to a spin-2 field propagating in a general curved spacetime. We
proceed as follows. First consider the Bel-Robinson tensor:
\be
T^{\mu\nu\rho\sigma} = 4\sigma^{\mu}_{AA^\prime}
\sigma^{\nu}_{BB^\prime}\sigma^{\rho}_{CC^\prime}
\sigma^{\sigma}_{DD^\prime}\psi^{ABCD} 
{\bar\psi}^{A^\prime B^\prime C^\prime D^\prime}~.
\ee
In spherically symmetric spacetimes a special role is played by
the specific component:
\be
T^{rttt} = - {1\over 16\Sigma\Delta^3} (|P_2 S_2|^2 - 
|P_{-2}S_{-2}|^2)
\ee
The energy momentum tensor for weak gravitational field can be
formed using the timelike Killing vector $\partial_t$:
\be
{1\over 4\pi}{dE\over dt} = - 
\int T^{rttt}g_{tt} ({\Delta\over\Sigma\omega})^2~
{\Sigma d\Omega\over 4\pi}
= {1\over 16\Sigma^3\omega^2}(|P_2|^2 - |P_{-2}|^2)
\label{eq:gravenergy}
\ee
This expression is applicable only in the weak field case, { i.e.} 
in the asymptotic regime. If it were rewritten in terms of the metric 
tensor it would be related to the second 
time derivative $\sim {\ddot h}^2$. This should contrasted with, {\it e.g.},
electromagnetic waves where only a single time derivative appears.
The extra derivative was compensated for by dividing out
with the square of the frequency, thus arriving at an expression
with correct dimensions.
 
The energy momentum inferred from the Bel-Robinson tensor is conserved
in all static black hole spacetimes. However, in order to arrive at a 
proper measure of 
graviton number close to the horizon we must integrate with respect to 
proper distance along the world line, rather than simply dividing
by the redshifted frequency. The resulting energy flux close to the horizon 
is given by eq.~\ref{eq:gravenergy} with the replacement 
$\omega^2\rightarrow \omega^2 + ({2\pi\over\beta_H})^2$. The flux
factor now becomes:
\be
|T_{s=2}|^2 =  {1\over 64\Sigma^3_{\rm hor}\omega^4}~{1\over 
\omega^2+ ({2\pi\over\beta_H})^2}~
| { A^{(0)}_2\over A^{(\infty)}_2 }|^2 ~,
\label{eq:s2flux}
\ee
where $\Sigma_{\rm hor} = {1\over 8\pi}\mu\beta_H$. The peculiar
polynomial dependence in $\beta_H$ is novel, and destined for a special
role in the argument.

In the book by 
Chandrasekhar~\cite{chandra} a different procedure is employed to find 
the flux factor for a spin-2 field. The starting point is the first law 
of thermodynamics that relates changes of the black hole mass 
to changes of  the horizon area. Subsequently variations in the area are 
related
to variations in the spin--coefficients, using a focusing theorem (essentially 
eq.~\ref{eq:psi0}). Finally the variations in spin-coefficients are
identified with the perturbation in the Weyl tensor, using the Ricci
identities. We have adapted this procedure to our case and recovered the 
result eq.~\ref{eq:s2flux}. This gives confidence that the flux factor 
for the graviton has been correctly identified.

\section{The Absorption Cross-section}
\label{sec:abs}
The general field equations cannot be solved exactly. However, we are 
interested in various limits where the absorption is weak. These cases 
have the dual advantage that analytical approximations are available,
and that the results have the same form as correlation functions in 
string theory. The limits are:
\begin{itemize}
\item
The dilute gas regime: exactly three of the four boost parameters 
are large, say $\delta_i\gg 1~;i=1,2,3$ with $\delta_4\sim 1$. 
In this case the black hole parameters are such that 
the inverse temperatures $\beta_R\sim\beta_L\sim\beta_H$ are much larger than 
the size of the black hole.
\item
Very low energy perturbations, $\omega\rightarrow 0$. The black hole
parameters can be arbitrary.
\item
Large partial wave number $j\gg \sqrt{\mu\omega}$. The black hole 
parameters can be
arbitrary.
\end{itemize}
Either of these limits is sufficient to allow the approximation
scheme that has recently been exploited in similar calculations (references
include~\cite{wadia96,greybody,klebanov96b,mathur97,strominger97a,cl97a}):
first we solve the equations in the vicinity of the black hole, 
then in the region far from the black hole, and finally the solutions 
are matched in the intermediate region. A particular point is that 
some results apply to arbitrary black holes and so suggest that the 
effective string model applies even far from extremality~\cite{fl97,cl97a}.

\paragraph{The horizon region:}
To find an equation that is accurate in the region close to the
horizons we expand the function $\Sigma$ defined in eq.~\ref{eq:Sigma}: 
\bea
\Sigma^2 &\simeq & 
{1\over 8}\mu^3 (r+{1\over 4}\mu)\prod_i \cosh^2\delta_i 
-{1\over 8}\mu^3 (r-{1\over 4}\mu)\prod_i \sinh^2\delta_i 
+{\cal O}(r^2 - {1\over 16}\mu^2) \\
\Sigma &\simeq & {1\over 2}\mu (r+{1\over 4}\mu)\prod_i \cosh\delta_i 
-{1\over 2}\mu (r-{1\over 4}\mu) \prod_i \sinh\delta_i 
+{\cal O}(r^2 - {1\over 16}\mu^2)~.
\eea
In the first equation the omitted term is a polynomial with
an explicit factor of $r^2 - {1\over 16}\mu^2$ while in the
second it is an irrational function that vanishes at both
the horizons. Introducing the surface accelerations at the outer
and inner horizon:
\be
{1\over\kappa_{+}} = \mu\prod_i \cosh\delta_i~~~;~~
{1\over\kappa_{-}} = \mu\prod_i \sinh\delta_i~,
\ee
and the radial variable $x$, defined through $r={1\over 2}\mu x$,
we find the field equation that applies in the horizon
region:
\bea
&~& \{
(x^2-{1\over 4})^{s}
{\partial\over\partial x}(x^2 -{1\over 4})^{1-s}
{\partial\over\partial x}
-(j+s)(j+1-s)+ \nonumber \\ &+&
{1\over x-{1\over 2}}~[
({\omega\over 2\kappa_{+}})^2-{is\omega\over 2\kappa_{+}}]-
{1\over x+{1\over 2}}~[
({\omega\over 2\kappa_{-}})^2+{is\omega\over 2\kappa_{-}}] \}
P_{s}^{(0)}= 0~.
\eea
We consider the equation for the upper component of the wave function, 
with the one for the lower component found by complex conjugation.
This field equation is a second order differential equation and so
has two linearly independent solutions. A basis can be chosen so that 
they satisfy ingoing and outgoing 
boundary conditions at the horizon, respectively. In the
absorption geometry it is the ingoing solution that is
relevant:
\bea
P_{s}^{(0)}(x) &=& A^{(0)}_s ({x-{1\over 2}\over 
x+{1\over 2}})^{-i{\beta_H\omega\over 4\pi}}
(x+{1\over 2})^{s-1-j}\times \nonumber \\
&\times & 
F(1+j-i{\beta_R\omega\over 4\pi}, 1+ j -s - i{\beta_L\omega\over 4\pi},
1-s-i{\beta_H\omega\over 2\pi}, {x-{1\over 2}\over x+{1\over 2}} )~,
\label{eq:horsol}
\eea
where $F$ is the hypergeometric function. 
At large $x\gg 1$ the last argument of the hypergeometric function 
approaches the radius of convergence $|z|=1$ so this representation 
of the wave function is inappropriate for asymptotic expansion at
large $x$. This situation
can be rectified by a modular transformation of the hypergeometric
function\footnote{Note that in the hypergeometric function
of eq.~\ref{eq:horsol} the sum of the first two arguments minus the 
third equals an integer. This is a degenerate case that needs 
special considerations, as given in, {\it e.g.}, ~\cite{abramowitz}. 
The appearance
of logarithmic terms is related to this special circumstance.}.
In the new representation:
\be
P_{s}^{(0)} \sim A_s^{(0)} x^{s+j}
{ \Gamma( 2j+1 )\Gamma(1 - s - i{\beta_H\omega\over 2\pi})
\over \Gamma ( 1 + j - i{\beta_R\omega\over 4\pi})
\Gamma ( 1 + j - s - i{\beta_L\omega\over 4\pi})}
[ 1 + {\cal O} ( {1\over x})
+ {\cal O} ( {\log x\over x^{2j+1}} )
]~.
\label{eq:largexhor}
\ee
The leading power corrections and the leading logarithmic 
corrections have been indicated independently. 

\paragraph{The asymptotic region:}
The next step is to expand for large $r$. Terms of order $r^2$ and $r$ 
are retained, as is the angular momentum eigenvalue. After introducing 
the rescaled variable $x=2r/\mu$ that was also used in the horizon 
region the equation can be written:
\be
[x^{2s} \partial_x x^{2(1-s)}\partial_x
+{1\over 4}\mu^2\omega^2 x^2 + {1\over 4}\mu M\omega^2 x + is\mu\omega x
-(j+s)(j+1-s)]P_{s}^{(\infty)} = 0
\ee
The regular solution is proportional to:
\bea
P_{s}^{(\infty)} &=& A^{(\infty)}_{s}~{\Gamma (1+j+s+{i\over 4}M\omega)
\over \Gamma(2j+2)}~ 
(\mu\omega x)^{j+s}~e^{-{i\over 2}\mu\omega x+{1\over 8}\pi M\omega}
\times \nonumber \\
&\times & M_K (1+j-s-{i\over 4}M\omega, 2j+2, i\mu\omega x)~.
\eea
With the overall normalization that has been indicated the asymptotic 
behavior at large $x$ is:
\be
P_{s}^{(\infty)} \simeq  A^{(\infty)}_{s} 
(\mu\omega x)^{2s-1}~e^{-{i\over 2}\mu\omega x+
{i\pi\over 2}(1+j-s)}~[1 + {\cal O} ({1\over x})] ~,
\ee
while at small $x$:
\be
P_{s}^{(\infty)} \simeq  A^{(\infty)}_{s}~{e^{{\pi\over 8}M\omega}
\Gamma (1+j+s+{i\over 4}M\omega)
\over \Gamma(2j+2)} (\mu\omega x)^{j+s}~[1 + {\cal O}(x)]~.
\label{eq:smallxas}
\ee

\paragraph{The matching:}
At this point we combine the approximate solutions that apply in 
separate regions and form a single wave function that can be used
throughout. To justify this procedure in the case of particles
with spin we must extend the arguments previously given for scalar
particles. The wave equation includes a term of the schematic form:
\be
{\Sigma^2 \omega^2\over \Delta}
\sim x^2 \mu^2 \omega^2 + x\mu M\omega^2 - \mu^2 \omega^2 e^{4\delta}
+{1\over x} \beta_R \beta_L \omega^2~.
\label{eq:sgmtmp}
\ee
In the dilute gas regime the parameters of the black holes are
$M\sim\mu e^{2\delta}$ and $\beta_R\sim\beta_L\sim \mu e^{3\delta}$;
and frequencies satisfy $\mu\omega\ll e^{-2\delta}$. 
The restriction on frequency is mild enough that the interesting 
region $\beta_R\omega \sim \beta_L\omega \sim 1$ is covered.
Now, for $x\sim e^{2\delta}$ we find by inspection of eq.~\ref{eq:sgmtmp}
that {\it all terms are small}. Similarly, in the same range of $x$:
\be
{\omega\Sigma\partial_r\Delta\over \Delta} =
{\omega\Sigma\over\sqrt{\Delta}}~ {\partial_r \Delta\over\sqrt{\Delta}}
\sim {\omega\Sigma\over\sqrt{\Delta}}
\ll 1
\ee
and, after a short calculation:
\be
\omega s\partial_r\Sigma\sim\mu\omega e^{2\delta} \ll 1~.
\ee
Thus the radial dependence of the wave function is determined by the 
kinetic term for $x$ in this range. At smaller $x$ the horizon
approximation gives the dominant terms, at larger $x$ the 
asymptotic equation is accurate, and in the matching region both
equations apply and are dominated by the kinetic operator. It
is therefore justified, in the dilute gas regime, to identify
the large $x$ approximation of the horizon wave function 
(eq.~\ref{eq:largexhor}) with the
small $x$ limit of the asymptotic wave function 
(eq.~\ref{eq:smallxas}).

When the frequency is {\it very} small $\mu\omega\rightarrow 0$ the 
matching procedure is similarly justified: the matching region can be chosen 
at $x\sim 1$, while avoiding the outer horizon at $x={1\over 2}$.
For large partial wave number $j\gg (\mu\omega)^{1/2}$ the argument
is slightly different~\cite{cl97a,cl97b}: in this case there exist
a matching region $x\sim 1$ where the angular momentum term $\sim j^2$
{\it dominates all other terms}. In this region the wave function 
is determined by the kinetic term and the angular momentum term while, 
at $x$ smaller or larger than the matching region, the horizon and the 
asymptotic equation applies, respectively. 

Applying the matching procedure we find:
\bea
| { A^{(0)}_{s}\over A^{(\infty)}_{s} }|^2 &=& 
(\mu\omega)^{2j+2s} ~
{e^{{\pi\over 4}M\omega}
|\Gamma (1+j+s+{i\over 4}M\omega)|^2\over \Gamma(2j+2)^2\Gamma(2j+1)^2}\times
\nonumber \\
&\times &|{\Gamma (1+j+i{\beta_R\omega\over 4\pi})
\Gamma (1+j-s+i{\beta_L\omega\over 4\pi})
\over\Gamma (1-s+i{\beta_H\omega\over 2\pi})} |^2~.
\label{eq:arat}
\eea
Taking into account the flux factors, considered in the previous
section, this translates into the absorption cross-section:
\bea
\sigma_{\rm abs}(\omega) &=& \omega^{2j-1}\mu^{2j+1}
~{e^{{\pi\over 4}M\omega}
|\Gamma (1+j+s+{i\over 4}M\omega)|^2\over \Gamma(2j+2)^2\Gamma(2j+1)^2}
(2j+1)\times  \nonumber  \\
&\times &|\Gamma (1+j+i{\beta_R\omega\over 4\pi})
\Gamma (1+j-s+i{\beta_L\omega\over 4\pi})|^2\times
\left\{ \begin{array}{cl}
{\pi\over |\Gamma( {1\over 2} + {i\beta_H\omega\over 2\pi})|^2}
\\
{\beta_H\omega\over 2|\Gamma( 1 + {i\beta_H\omega\over 2\pi})|^2} 
\end{array} \right.
\label{eq:abscs}\eea
where the upper line is for half-integer spin $s={1\over 2}$
and the lower line is for integer spin $s=0,1,2$. 
At this point we recall the identities:
\be
e^{-{\alpha\over 2}}|\Gamma ({1\over 2}+i{\alpha\over 2\pi}) |^2
={2\pi\over e^\alpha+1}~~~;~~
e^{-{\alpha\over 2}}|\Gamma (1+i{\alpha\over 2\pi}) |^2
={\alpha\over e^\alpha-1}~.
\ee
The last terms in the absorption cross-section therefore
gives rise precisely to the thermal factors {\it in the numerator}
that are needed to cancel the thermal factors 
{\it in the denominator} of Hawking's expressions eq.~\ref{eq:hawb}
and eq.~\ref{eq:hawf} for the emission rates of bosons and fermions,
respectively. Thus the emission rate becomes:
\bea
\Gamma_{\rm em}(\omega)
&=& \omega^{2j-1}\mu^{2j+1}
~{e^{{\pi\over 4}M\omega}|
\Gamma (1+j+s+{i\over 4}M\omega)|^2\over 2\Gamma(2j+2)^2\Gamma(2j+1)^2}
(2j+1)\times \nonumber \\
&\times&e^{-{\beta_R\omega\over 4}}
|\Gamma (1+j+i{\beta_R\omega\over 4\pi})|^2~
e^{-{\beta_L\omega\over 4}}
|\Gamma (1+j-s+i{\beta_L\omega\over 4\pi})|^2~{d^3 k\over (2\pi)^3}
\label{mr}\eea
This expression is our main result. It exhibits interesting
features: the dependence on Hawking 
temperature disappeared and so the factorized form expected from an 
effective string description is manifest. Moreover, the spin
dependence takes a surprisingly simple form. 

In eq.~\ref{eq:arat} for the ratio of the wave function 
normalizations at the horizon and at infinity a factor of 
$|\Gamma(1-s+i{\beta_H\omega\over 2\pi})|^2$ appears in the denominator. 
This is a potential source 
of problems for  $s>{1\over 2}$ because, after expansion of the Gamma-function,
there will be a polynomial dependence on $\beta_H$ in addition
to the thermal factors. It would be very difficult to reconcile
a dependence on $\beta_H={1\over 2}(\beta_R+\beta_L)$ with the 
factorized correlation functions expected from an effective string 
description. It is therefore  important that the flux factors 
derived in the previous section precisely cancel these potentially dangerous
 terms.
The cancellation that appears here is a novel one that does not show up 
when considering spin-0 particles.

The non-extremality parameter $\mu$ parameterizes the coordinate distance 
between the two horizons $\mu = 2(r_{+} - r_{-})$. This quantity therefore 
does not appear to have fundamental significance. In macroscopic
applications it is natural to trade it for the entropy $S$ and $\beta_H$:
\be
S= 2A ={\mu\beta_H}~.
\label{eq:amu}
\ee
It is less clear what the appropriate microscopic variable is, but
a natural representation is to eliminate $\mu$ in favor of the effective 
string length ${\cal L}$ given by~\cite{cl97a,kastor}:
\be
{\cal L} = {\mu\beta_R\beta_L\over 2\pi}= 
2\pi\mu^3 (\prod^{3}_{i=1}\cosh^2\delta_i-
\prod^3_{i=1}\sinh^2\delta_i) ~.
\label{eq:sl}
\ee

\section{Results}
\label{sec:results}
In this section we write the results  of the previous section 
in a more digested form and discuss their structure.

\paragraph{Large angular momentum:} In this case  ($j>> \sqrt{\mu\omega}$)
the  expression for the   emission rate  
(eq. \ref{mr})  is valid for arbitrary black hole background. The result can
be rewritten  explicitly in terms of the thermal factors as follows:
\bea
\Gamma_{\rm em}^{\rm boson}(\omega)&=&A_{j} {\cal C}_{j+s}
{P_R({\omega\over 2})\over e^{\beta_R\omega/2}-1}~
{P_L({\omega\over 2})\over e^{\beta_L\omega/2}-1}~{d^3 k\over (2\pi)^3}~,
\label {eq:hawbp}
\\
\Gamma_{\rm em}^{\rm fermions}(\omega)&=&A_{j}{\cal C}_{j+s}
{P_R({\omega\over 2})\over e^{\beta_R\omega/2}+1}~
{P_L({\omega\over 2})\over e^{\beta_L\omega/2}-1}~{d^3 k\over (2\pi)^3}~.
\label {eq:hawfp}
\eea
The $A_j$ are the prefactors:
\be
A_{j}={{\omega^{2j-1}\mu^{2j+1}(2j+1)}\over{2[(2j+1)!(2j)!]^2}},
\label {eq:aj}\ee
that are independent of the mass $M$ and the spin $s$.
This dependence is carried by the Coulomb factors ${\cal C}_{j+s}$ 
given by:
\be
{\cal C}_{j+s} ={P_C({\omega\over 2})\over {1-e^{-\pi M \omega/2}}}, 
\label{eq:cjs}\ee
where:
\be
P_C({\omega\over 2})=\pi\omega{({M\over{2}})^{2j+2s+1}\prod_{k=1}^{j+s} [
({\omega\over 2})^2 + ({{2 k}\over M}})^2 ].
\ee
Note that ${\cal C}_{j+s}\rightarrow (j+s)!^2$ in the limit
$M\omega\rightarrow 0$. The exponential factors with parameter 
${1\over 4}\pi M=2\pi G_N M$ reflect the long range
interaction caused by gravity in four dimensions. These terms
are not specific to the black hole geometry, but it is nevertheless
interesting that their form is reminiscent of the thermal factors. This 
result suggests that Coulomb factors may also have a statistical origin in 
the microscopic theory. 
 
The polynomials $P_{R}({\omega\over 2})$ are: 
\begin{eqnarray}
P_{R}^{\rm bosons}({\omega\over 2})
&=&\pi \omega ({\beta_R\over 
{2\pi}})^{2j+1}\prod_{k=1}^{j} [({\omega\over 2})^2
+ ({2\pi k\over\beta_R})^2 ]~,\\ 
P_{R}^{\rm fermions}({\omega\over 2})
&=& 2\pi ({\beta_R\over {2\pi}})^{2[j]+2}
\prod_{k=1}^{[j]+1} [({\omega\over 2})^2 
+ ({\pi (2k-1)\over\beta_R})^2 ]~,
\label{eq:pr}
\end{eqnarray}
for bosons and fermions, respectively, and the $P_L({\omega\over 2})$ are:
 \be
P_{L}({\omega\over 2})=\pi \omega ({\beta_L\over{2\pi}})^{2j-2s+1}\prod_{k=1}^{
j-s} [
({\omega\over 2})^2 + ({2\pi k\over\beta_L})^2 ].
\label{pl}
\ee
for both fermions and bosons. The emission rates 
eqs.~\ref{eq:hawbp}-\ref{eq:hawfp} has precisely the 
factorized form that was advertised in  the introduction 
(eq.\ref{eq:hawb} and eq.~\ref{eq:hawf}). 
For scalars the result $s=0$ agrees  with that of~\cite{mathur97,cl97b}. 
Interestingly, the thermal factor of the left-moving sector is always of 
the Bose-Einstein type, but the thermal factors of the right-moving sector
correlate with the spin-statistics of the emitted particles: they are of 
the Bose-Einstein type for bosons and of the Fermi-Dirac type for fermions.
This result is in agreement with the structure of the thermal
quantities of rotating four-dimensional black holes: only the right-moving 
temperature and entropy depend on the angular momentum of the black hole.

\paragraph{Minimal total angular momentum:}
In this case $j=s$  and the emission rates
eqs.~\ref{eq:hawbp}-\ref{eq:hawfp}  
are valid for the  black hole   parameters of the dilute
gas regime. Then the rates take a relatively simple form:
\bea
\Gamma_{\rm em}^{s=0}(\omega)&=&
{1\over 2\omega}\mu
{{{\beta_R\omega}/2}\over e^{\beta_R\omega/2}-1}~
{{{\beta_L\omega}/2}\over e^{\beta_L\omega/2}-1}~{d^3 k\over (2\pi)^3}~,\\
\Gamma_{\rm em}^{s={1\over 2}}(\omega)
&=&{\textstyle{1\over 4}}\mu^2
{{{\beta_R^2\over{2\pi}}[({\omega\over 2})^2 
+ ({\pi\over\beta_R})^2 ]}\over e^{\beta_R\omega/2}+1}~
{{{\beta_L\omega}/2}\over e^{\beta_L\omega/2}-1}~{d^3 k\over (2\pi)^3}~,
\\
\Gamma_{\rm em}^{s=1}(\omega)
&=&{\textstyle{1\over 24}}\omega \mu^3
{{\pi\omega ({\beta_R\over {2\pi}})^{3}
[({\omega\over 2})^2 
+ ({2\pi \over\beta_R})^2 ]}\over e^{\beta_R\omega/2}-1}~
{{{\beta_L\omega}/2}\over e^{\beta_L\omega/2}-1}~{d^3 k\over (2\pi)^3}~,\\
\Gamma_{\rm em}^{s=2}(\omega)
&=&{\omega^3 \mu^5\over {10 (4!)^2}}
{{\pi\omega ({\beta_R\over {2\pi}})^{5}
\prod_{k=1}^2[({\omega\over 2})^2 
+ ({2\pi k\over\beta_R})^2 ]}\over e^{\beta_R\omega/2}-1}~
{{{\beta_L\omega}/2}\over e^{\beta_L\omega/2}-1}~{d^3 k\over (2\pi)^3}~.
\label {eq:jeshaw}
\eea
To arrive at these formulae we used the low energy form
of the Coulomb factor ${\cal C}_{2s}=[(2s)!]^2$.
In each case the left-moving sector  reduces to a single term thus
indicating a ``minimal'' (bosonic) excitation in this sector, 
while the right-moving sector  indicates progressively
 more involved structure of excitations as the 
spin of the particle increases. The {\it spin-dependence} 
of the emission
rates  is ``stored'' {\it  in  the right-moving sector}, only. 

\paragraph{The low-energy absorption cross-section:}
At very low energy the absorption is dominated by the leading
partial wave $j=s$ and the expression  eq. \ref{eq:abscs}
is valid in any black hole background. We find:
\bea
\sigma_{\rm abs}^{s=0}(\omega\rightarrow 0) &=& {1\over 2}\mu\beta_H = A~, \\
\sigma_{\rm abs}^{s={1\over 2}}(\omega\rightarrow 0) 
&=& {\pi\mu^2\over 8}~, \\
\sigma_{\rm abs}^{s=1}(\omega\rightarrow 0) &=& 
{\beta_H\omega^2\mu^3\over 24}~,\\
\sigma_{\rm abs}^{s=2}(\omega\rightarrow 0) &=& 
{\beta_H\omega^4\mu^5\over 1440}~.
\eea
The result for the minimally coupled fermions agrees with the expression
found in~\cite{gibbons96}\footnote{The $g_H$ of~\cite{gibbons96} is
$g_H = {8\beta_H^2\over 2\pi A}=16\prod_i\cosh\delta_i$.}. As noted in
that reference the absorption cross-section of minimally coupled 
fermions vanish in the extremal limit $\mu\rightarrow 0$.  
The result is also in agreement with the low energy absorption cross-section
for  the Schwarzschild black hole backgrounds  with  minimally coupled 
scalars and Dirac fermions~\cite{abs2}.

\section{Microscopic Interpretation}\label{sec:microscopics}
The factorized form of the emission rates eq. \ref{mr} 
 indicates an interpretation in
terms of  multi-body annihilation rates in an effective string
model. We can write the  result for the emission rate 
as:
\be
\Gamma_{\rm em}(\omega) 
=A_j{\cal C}_{j+s} 
 \Gamma(2h_R)\Gamma(2h_L)({\beta_R\over {2\pi}})^{2h_R-1}
 ({\beta_L\over {2\pi}})^{2h_L-1}
 G^{h_R}_{\beta_R}\left({\omega\over 2}\right)
 ~G^{h_L}_{\beta_L}\left({\omega\over 2}
\right)~,
\label{eq:cftc}\ee
where:
\be
G^h_{\beta}\left({\omega\over 2}\right)\equiv ({{2\pi}\over{ \beta}})^{2h-1}
e^{-{\beta\omega\over 4}}  
{|\Gamma (h+{{i\beta\omega}\over 4\pi })|^2\over \Gamma(2h)}.
\label{eq:green}
\ee
The prefactor $A_j$ and  the Coulomb factor ${\cal C}_{j+s}$ are defined  in
eq.~\ref{eq:aj} and eq.~\ref{eq:cjs}, respectively.
In the effective string  description 
the $G^{h_{R,L}}_{\beta_{R,L}}({\omega\over 2})$ are the  
Fourier space representations of the canonically normalized thermal 
Green's functions for a primary field with the respective  right- and 
left-moving conformal dimensions $h_{R,L}$~:
\be
G^{h_R}_{\beta_R} (z) = \left({{\pi\over{\beta_R}} 
\over{\sinh ({{\pi z}\over \beta_R})}}\right)^{2h_R}, \ \ \ 
G^{h_L}_{\beta_L} ({\bar z}) = \left({{\pi\over {\beta_L}}
\over\sinh ({{\pi{\bar  z}}\over\beta_L})}\right)^{2h_L}~.
\ee
We now arrive at a qualitative interpretation of the emission in terms of 
a $(4,0)$ super-conformal field theory. The fields responsible for the
emission have conformal dimensions:
\be 
(h_R,h_L)=(j+1 ,j-s+1). 
\ee
The left-moving sector involves only fields with the integer
conformal dimensions (bosonic  string excitations), while the right-moving 
sector
involves conformal fields with  both  integer and   half-integer
conformal dimensions (both bosonic and fermionic string excitations). 
The effective string description is therefore ``heterotic''.
Note that we are describing the emission in terms of irrelevant
operators, i.e. $(h_R, h_L)\ge 1$. This result embodies the correct physics:
the amplitudes vanish for low energies with the power of frequency related 
to the conformal weight in a simple way.

A few remarks are in order:
\begin{itemize}
\item
Particles with the minimal total angular momentum $j=s$ correspond
to the conformal dimensions:
\be 
(h_R,h_L)=(1+s~,1)~.
\ee
We may interpret  the result as  the emission of 1 boson and 
$2s$ fermions
in the right-moving sector, colliding with 1 boson in the left-moving sector.
 The  pattern  of conformal weights also 
suggests that the emission rates of the 
{\it  nearby integer and half-integer spin} 
particles  are correlated in the microscopic picture with the
emission of the  {\it  nearby  components of the   world-sheet
superfields of the right-moving sector}~\footnote{Conformal dimensions of the 
components  of the   world-sheet superfield differ by 1/2.}.
It is possible  that the super conformal symmetry 
of the right-moving sector would enable one 
to derive a precise relationship between emission rates for particles with 
different spins. 
\item
The frequency dependence of the emission rates can be understood as 
follows. First of all the many-body kinematics of the initial state is 
taken into account by the Green's function. The prefactor scales as
$\omega^{2j-1}$ and is interpreted in the same way as the higher partial 
waves of minimally coupled scalar 
fields~\cite{strominger97a,gubser,mathur97b,cl97b}:
the outgoing wave has a  normalization $\omega^{-1}$ and each unit of angular 
momentum requires one derivative in the interaction term.
This gives $\omega^j$ in the amplitude, and $\omega^{2j}$ in the rate.
The same argument also applies to particles with spin, except that for 
fermions the wave function normalization is independent of $\omega$, and 
only the integer part of the angular momentum gives rise to a derivative 
term. 
\item
The dependence of the remaining dimensionful quantities is not
understood very well. A particular feature of the present problem is that 
the powers of $\beta_R$ and $\beta_L$  that appear in eq.~\ref{eq:cftc}
differ by a factor $2(h_R-h_L)=2s$.  This asymmetry in the couplings is 
likely to be related to the asymmetry between the two sectors. It is an 
important problem to construct a concrete model that exhibits this feature. 
This asymmetry  also prevents us from combining the 
$\beta_{R,L}$ and the $\mu$ parameter into a single parameter ${\cal L}$ 
--- the string length (eq. \ref{eq:sl}). In general, there are therefore
 different relevant scales.
\item
A microscopic calculation of the overall numerical coefficient 
in eq.~\ref{eq:cftc} is not  feasible, yet. However, the microscopic picture
that is needed should be  robust and so one expects
that  such a factor could be derived from 
group theoretic factors that could follow
directly from the algebra of the microscopic theory.
\end{itemize}
The  quantitative 
microscopic description of the effective string is thus incomplete.
However  the qualitative picture in terms of correlation functions of 
primary fields of a $(4,0)$ conformal field theory is already apparent 
at the current level of our understanding. In the recent microscopic
derivation of the entropy of extremal black holes in four dimensions
an auxiliary $(4,0)$ conformal field theory appears 
prominently~\cite{malda97,vafa97}. This is very suggestive, but the 
precise connection is not clear.

\section{Concluding remarks}
\label{sec:conclusion}
 We would like to conclude with a number of outstanding questions.
 
\begin{itemize}
 \item
It is an important task to determine the generality of the kind 
of greybody factors discussed in this paper. For example,
it is not yet clear if the special properties of black holes that can be 
interpreted as solutions of the effective toroidally compactified string 
theory, i.e. solutions to  specific  $N=8$ or $N=4$ supergravity are needed. 
\item
We did not complete the decoupling of the gravitational perturbation 
equations. It is actually not clear that this is possible; indeed it 
was a surprise when it was first accomplished for the Reissner-Nordstr\"{o}m
case. 
It would be desirable if the underlying supersymmetry supplies
the structure needed to ensure such a  decoupling.
\item
The Newman-Penrose formalism is peculiarly well-adapted to
the problem of black hole perturbations but it seems specific to four 
dimensions. It would be interesting to consider higher spins also
in the five-dimensional case. For the specific example of spin-1/2
fermions  some results were obtained in~\cite{hoso}. 
\item  Another major challenge is 
to allow for {\it rotating}  black hole
 backgrounds. In this case it is not obvious that 
the variables  in the field equations can be separated
thereby  reducing the problem to effective one dimensional scattering. 
It is indeed surprising that, for  minimally coupled scalar 
fields in rotating backgrounds, the equations do exhibit the desired 
separability both for four-dimensional~\cite{cl97b} and five-dimensional
~\cite{cl97a} general rotating black hole backgrounds. 
If a similar  simplification takes place 
for fields with spin the resulting greybody factors would necessarily 
be of the  form considered in this paper and the main
qualitative change that would be expected is the replacement:
\be
\beta_R\omega \rightarrow \beta_R\omega - m\beta_H\Omega~,
\ee
where $\Omega$ is the rotational velocity of the black hole and
$m$ is the projection of the total angular momentum onto the
axis of rotation. The  thermodynamic quantities   $\beta_{R,S}$ and $S_{R,L}$ 
would also acquire 
dependence on the rotational  parameter of the background, as  it was made 
explicit in~\cite{cl97b}. The reason that it is possible to anticipate 
the answer in the rotating case on general grounds is that the
computations leading to greybody factors of particular interest in 
string theory are independent of many details: they only 
depend on the structure of the black hole in the vicinity of the two 
horizons and at asymptotically large distances; and these features can be 
inferred from general principles (global spacetime structure of the
such black holes), rather than explicit calculations. 
The
robustness that manifests itself in this way suggests universal properties 
of the underlying microscopic theory. Specifically it would seem that 
the structure involving independent potentials of left and right moving 
modes should be  generic.
\end{itemize}

{\bf Acknowledgments:}
We would like to thank C. Callan, G. Gibbons, S. Mathur, K. Stelle,
and R. Wald for discussions. FL would like to thank the University
of Michigan at Ann Arbor for hospitality during part of this work;
and MIT for hospitality in its final stage. This work was supported 
by DOE grant DE-FG02-95ER40893.


\end{document}